\RequirePackage{fix-cm}
\documentclass[smallextended]{svjour3}       
\smartqed  
\usepackage [latin1]{inputenc}
\usepackage{graphicx}
\usepackage{dcolumn}
\usepackage{bm}
\usepackage{amsmath}
\usepackage{enumerate}
\usepackage{setspace}
\usepackage{dsfont}
\usepackage{amsfonts}

\begin{document}

\title{The $m$-least significant bits operation for quantum random number generation
}


\author{Ziyang Chen$^{1}$         \and
        Zhengyu Li$^{1}$  \and
        Bingjie Xu$^{2}$    \and
        Yichen Zhang$^{3}$      \and
        Hong Guo$^{1}$ 
}


\institute{
              \email{Hong Guo \\hongguo@pku.edu.cn}             \\
                        $1$ State Key Laboratory of Advanced Optical Communication, Systems and Networks, Department of Electronics, and Center for Quantum Information Technology, Peking University, Beijing 100871, China \\
                        $2$ Science and Technology on Security Communication Laboratory, Institute of Southwestern Communication, Chengdu 610041, China\\
                        $3$ State Key Laboratory of Information Photonics and Optical Communications, Beijing University of Posts and Telecommunications, Beijing 100876, China\\
}

\date{Received: date / Accepted: date}

\maketitle

\begin{abstract}
Quantum random number generators (QRNGs) can provide genuine randomness based on the inherent unpredictable nature of quantum physics. The extracted randomness relies not only on the physical parts of the QRNG, such as the entropy source and the measurement device, but also on appropriate postprocessing method. The $m$-least significant bits ($m$-LSBs) operation is one of the simplest randomness extraction method, which has the advantage of easy implementations. Nonetheless, a detailed analysis of the $m$-LSBs operation in QRNGs is still missing. In this work we give a physical explanation of the $m$-LSBs operation by introducing a new positive operator-valued measurement operator, which is obtained by regrouping the results of coarse-grained measurements. Both trusted and untrusted source scenarios are discussed. The results show that the $m$-LSBs operation can extract randomness effectively under the condition of the trusted source, while it is not effective under the untrusted source scenario.
\end{abstract}

\section{Introduction}
\label{intro}
Random numbers are widely used in many fields, including numerical simulations \cite{J.Am.Stat.Assoc.44.335.1949}, statistical analysis \cite{Design.and.Analysis}, fundamental physics tests \cite{Physics.1.195.1964}, and cryptography \cite{Bell.System.Technical.Journal.28.656.1949}. Traditionally, pseudo-random number generators (PRNGs), which are based on deterministic computational algorithms aiming at expending a short random seed into a longer random sequence, are adopted to generate random bits in extensive applications. The randomness generated from PRNGs is determined by the short random seed, thus any random numbers generated by algorithms cannot be truly random even if the sequence appears ``random''. However in some fields like cryptography, where the unpredictable feature of the random numbers is of great importance, PRNGs cannot meet the requirements of implementations.

In contrast, quantum random number generators (QRNGs) \cite{npj.Quantum.Information.2.16021.2016,Rev.Mod.Phys.89.015004.2017,Rep.Prog.Phys.80.124001.2017} can provide information-theoretical provable random numbers, which are in principle unpredictable and irreproducible based on the intrinsic unpredictable nature of quantum physics, thus they have attracted much attentions in recent years. In general, a QRNG can be divided into four parts: the randomness source (also called the entropy source), the measurement device, the postprocessing process, and the randomness test. The process of generating quantum random numbers can be described as follows \cite{npj.Quantum.Information.2.16021.2016,Rev.Mod.Phys.89.015004.2017}: An entropy source is prepared in a superposition of the measurement states, and a measurement operator is performed on the prepared states. The measurement outcomes are intrinsically unpredictable according to Born's rule. However, a perfect state preparation with an ideal measurement is always a difficult task in experiment, leading to the generation of the random numbers not being able to reach the ideal case. Therefore, the postprocessing step needs to be performed.

According to the requirements of the physical devices, QRNGs can be classified into three different types \cite{npj.Quantum.Information.2.16021.2016}, namely, \emph{full-trusted-device}, \emph{semi-device-independent} and \emph{full-device-independent} QRNGs.

In trusted-device QRNG scenarios, schemes of various entropy sources have been demonstrated, including photon spatial modes \cite{Rev.Sci.Instrum.71.1675.2000,J.Mod.Opt.47.595.2000,J.Appl.Phys.100.056107.2006} and temporal modes \cite{Appl.Opt.44.36.7760.2005,Rev.Sci.Instrum.78.045104.2007,Appl.Phys.Lett.93.031109.2008,J.Mod.Opt.56.516.2009,Appl.Phys.Lett.98.171105.2011,Appl.Phys.Lett.104.051110.2014}, photon number distribution \cite{Opt.Lett.34.12.1876.2009,Phys.Rev.A.83.2.023820.2011}, phase noise of lasers \cite{Phys.Rev.E.81.5.051137.2010,Opt.Lett.35.3.312.2010,Opt.Express.19.21.20665.2011,Appl.Phys.Lett.104.26.2014,Opt.Express.20.12366.2012,Rev.Sci.Instrum.86.6.063105.2015,Phys.Rev.A.91.062316.2015,Opt.Express.24.24.27475.2016}, quadrature fluctuation of vacuum noise \cite{Nat.Photonics.4.711.2010,Phys.Rev.A81.063814.2010,Appl.Phys.Lett.98.231103.2011}, amplified spontaneous noise \cite{Opt.Express.18.23.23584.2010,Opt.Lett.36.6.1020.2011,IEEE.Photon.Technol.Lett.24.6.437.2012,Laser.Phys.Lett.10.045001.2013,IEEE.J.Lightwave.Technol.33.13.2855.2015,Rev.Sci.Instrum.88.113101.2017} and so forth, which assume that all the device models are fully characterized to pursue a high generation rate. However, in practical, the devices are always interacted with the environment, resulting in the hard calibrations of them. Even worse, the devices may leak side information to potential adversaries, leading to a better prediction of random numbers. This case is sometimes called \emph{untrusted-device} scenario.

To deal with the scenario of untrusted devices, self-testing QRNGs \cite{Phys.Rev.Lett.114.150501.2015}, e.g. device-independent (DI) QRNGs \cite{Nature.464.1021.2010,Nature.497.227.2013}, were proposed to generate genuine randomness without perfectly characterising the practical instruments, and therefore, the output randomness does not rely on any assumptions about the physical devices. Despite today's experimental techniques make the speed of DI-QRNGs a tremendous progress \cite{Phys.Rev.Lett.120.010503.2018,Phys.Rev.Lett.120.040406.2018,arXiv.1807.09611}, it is still hard to fulfil people's needs in many fields. As a compromise, the semi-device-independent QRNG offers a good trade-off between the performance of the trusted-device QRNG and the security of the DI-QRNG, thus this scheme has great potential. The secure random numbers can be generated by trusting partial devices of the QRNG, while the assumptions of other devices are removed. In general, there are two main branches of semi-device-independent QRNGs: one is the measurement-device-independent QRNG \cite{Rev.Applied.7.054018.2017,New.J.Phys.17.125011.2015,Phys.Rev.A.94.060301.2016}, and the alternative is the source-device-independent (SDI) QRNG \cite{Phys.Rev.A.90.052327.2014,Phys.Rev.X.6.011020.2016,Phy.Rev.Lett.118.060503.2017,Nat.Commun.9.5365.2018,arXiv.1801.04139.2018,arXiv.1709.00685.2017,Phys.Rev.A.99.022328.2019}.

Nevertheless, only good-quality entropy sources and well-characterized detectors are not enough for a QRNG. The postprocessing procedure is necessary due to the following two reasons: randomness extracted from the environment needs to be eliminated and uniform randomness from the randomness source needs to be extracted. The postprocessing process generally includes the entropy estimation phase and the randomness extraction phase \cite{Phys.Rev.A.87.062327.2013}. The entropy estimation phase tells us how many uniform random bits can be extracted at most in the raw data (the random sequences without postprocessing) with distribution $X$, namely the min-entropy ${H_\infty }\left( X \right)$ in trusted-device scenarios and the conditional min-entropy ${H_\infty }\left( {X{\rm{|}}E} \right)$ in untrusted-device scenarios. The randomness extraction step aims at using extractors to distill (almost) uniform randomness based on the entropy estimation phase.

There are two categories of extractors: deterministic extractors and seeded extractors. Deterministic extractors are functions taking input $n$-bits strings into shorter output $m$-bits strings. Some simple postprocessing methods have been widely used in QRNGs, such as the $m$-least significant bits ($m$-LSBs) operation \cite{Phys.Rev.E.81.5.051137.2010,Appl.Phys.Lett.98.231103.2011,Nat.Photonics.2.728.2008,Phys.Rev.Lett.103.24102.2009}, the logical exclusive-OR operation \cite{Opt.Lett.35.3.312.2010}, the von Neumann debiasing method \cite{Opt.Lett.34.12.1876.2009} and so forth. They are charming for the simple algorithms which are easy to implement in both hardware and software implementations, and thus suitable for the high-speed and real-time scenarios. They also have the advantage that the procedures do not need pre-prepared random seeds to work, which are sometimes limited resources in the real-life realizations. The alternative seeded extractors are functions inputing $n$-bits raw data and $k$-bits uniform random seeds and producing $m$-bits random sequences. The superiority of seeded extractors is that the extracted randomness of them can be proved as information-theoretical-provable randomness, e.g. Toeplitz-hashing \cite{Lecture.Notes.in.Computer.Science.Vol.893,Theor.Comput.Sci.107.235.2002} and Trevisan's \cite{J.ACM.48.2001.1999,31st.Annual.ACM} extractors. Nonetheless, even today's real-time applications have made great progress~\cite{arXiv:1812.05377.2018,Review.of.Scientific.Instruments.90.043105.2019}, the rather low speeds and the complex structures restrict the applications especially in the real-time implementations.

In this paper, we focus on the most simple extractors: deterministic extractors, especially on the $m$-LSBs operations, which are still lacking of a detailed analysis in QNRGs (which were only discussed in Ref.\cite{Appl.Phys.Lett.98.231103.2011} to our knowledge). We analyze the relationship between the data before and after the postprocessing procedure and then construct operators of the $m$-LSBs operations to avoid sophisticated mathematical discussions. The two main contributions of this work can be summarized as follows:

(1) We restrict our discussions on the \emph{coherent-detection} (homodyne or heterodyne detection) types of QRNGs, and build new positive operator-valued measurement (POVM) operators of the $m$-LSBs operations.

(2) We apply the POVM operators to analyse the performances of $m$-LSBs operators in both trusted source and untrusted source scenarios. The entropy estimation steps are performed by acting the new operators on quantum states in the trusted source scenario and constructing a new entropic uncertainty relation (EUR) in the untrusted source scenario, respectively.

The rest of this paper is organized as follows. In Sec.\ref{POVM_operator_of_LSB}, we build new POVM operators of the $m$-LSBs operation. Then, we analyse the performances of the $m$-LSBs operations in both trusted source scenario (Sec.\ref{Trusted_source_scenario}) and untrusted source scenario (Sec.\ref{Untrusted_source_scenario}), followed by a discussion in Sec.\ref{Discussion}. Finally, a summary of the paper is given in Sec.\ref{Conclusion}.

\section{The POVM operator of the $m$-LSBs operation}
\label{POVM_operator_of_LSB}
In this section, we restrict our discussion on the QRNG schemes with coherent detection, which are one of the promising continuous-variable QRNG techniques \cite{Rev.Mod.Phys.89.015004.2017} and they have the advantages of the high generation rates and easy to implement with ``off-the-shelf'' detectors. First, we review the coarse-grained position-momentum measurements introduced by the finite precision and finite range of detectors. Then we analyse the feature of the coarse-grained measurement intervals and recombine the measurement intervals according to the data's characteristics after the $m$-LSBs operation. Last but not least, we define new $m$-LSBs operators based on the recombined intervals.

\begin{figure}[b]
\centering
\includegraphics[width=8cm]{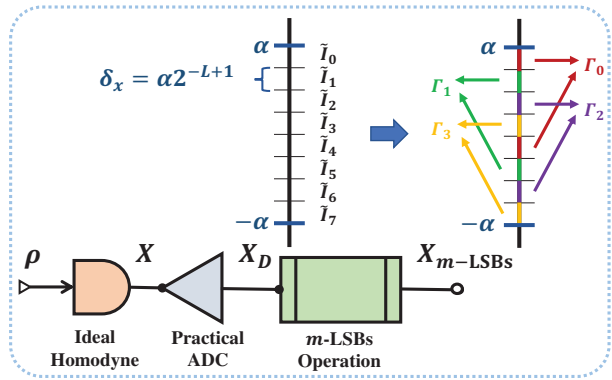}\\
\caption{(Color online) The general structure of the measurement and the $m$-LSBs postprocessing procedure in QRNGs. A practical homodyne detector is modelled by an ideal homodyne detector followed by a practical ADC with sampling range from $- \alpha$ to $\alpha$ and $L$ is the digitized bits. The data after the $m$-LSBs operation can be recombined into new intervals. Here we choose $L=3$ and $m=2$ as an example.}
\label{practical_detection}
\end{figure}

The general structure of the practical (homodyne) detection and the $m$-LSBs postprocessing operation in QRNGs are shown in Fig.\ref{practical_detection}. A practical homodyne detector is always modeled by an ideal homodyne detector followed by a practical analog-to-digital converter (ADC) with finite sampling range and resolution. In the most basic scenario, the randomness source $\rho$ is measured by an ideal homodyne detection to get results $X$, which can be described by the projection operators $\hat M_x$ and $\hat N_p$ ($x,p \in \left( { - \infty , + \infty } \right)$):
\begin{equation}\
\hat M_x = \left| x \right\rangle \left\langle x \right|,      \quad\quad    \hat N_p = \left| p \right\rangle \left\langle p \right|,
\label{X_measurement_operator}
\end{equation}
where $\int_x {\left| x \right\rangle \left\langle x \right|} dx = 1$ and $\int_p {\left| p \right\rangle \left\langle p \right|} dp = 1$ are the normalization conditions. After measuring the quantum state, the measurement results need to be discretized, since practical scenarios are always performed with finite precisions. Considering that a practical ADC with finite sampling range (assuming the sampling range is from $- \alpha$ to $\alpha$) and resolution (assuming the digitized bits is $L$) is used for discretizing quadrature $X$, and the precision is ${\delta _x} = \frac{{2\alpha }}{{{2^L}}} = \alpha {2^{ - L + 1}}$. The measurement outcomes can be grouped into intervals $\left\{ {{I_i}} \right\}$, given by \cite{Phys.Rev.Lett.109.100502.2012}
\begin{equation}\
\left\{ {\begin{array}{*{20}{c}}
{{I_i} = \left( { - \infty , - \alpha } \right]}&{i = {\rm{0}}}\\
{{I_i} = \left( { - \alpha  + \left( {i - 1} \right){\delta _x}, - \alpha  + i{\delta _x}} \right],}&{i \in \left[ {{\rm{1}},{2^L}} \right]}\\
{{I_i} = \left( {\alpha , + \infty } \right)}&{i = {2^L}{\rm{ + 1}}}
\end{array}} \right.
\end{equation}
 Therefore the ideal homodyne detection together with the discretization processes formally corresponds to ${2^L} + 2$ POVM operators $\left\{ {\hat M_{{\delta _x}}^i} \right\}$ with elements $\hat M_{{\delta _x}}^i = \int_{{I_i}} {dx} \left| x \right\rangle \langle x|$, $\left( {i \in \left[ {0,{2^L} + 1} \right]} \right)$. The outcome $X_D^i$ after the discretization process appears with the probability of $\Pr \left( {X_D^i} \right) = {\rm Tr}\left( {\rho \hat M_{{\delta _x}}^i} \right)$. Noting that in experiment the value of $X_D^0$ is always treated as the same value of $X_D^1$, and the value of $X_D^{2^L+1}$ is treated as the same value of $X_D^{2^L}$, since infinite intervals ${I_0}$ and ${I_{{2^L} + 1}}$ correspond to the minimum and maximum values of the measurement outcomes, respectively. The probabilities should verify the normalization condition, i.e., $\sum\nolimits_i {\Pr \left( {X_D^i} \right)}  = 1$.

Due to the fact that the intervals ${I_0}$ and ${I_{{2^L} + 1}}$ are with infinite lengths, and they do not correspond to new measured values in the measurement results, leading to different considerations in our following analysis under different scenarios. Therefore, we define new discrete intervals $\left\{ {{{\tilde I}_i}} \right\}$ with same interval lengths, which satisfy
\begin{equation}\
{\tilde I_i} = \left( { - \alpha  + i{\delta _x}, - \alpha  + \left( {i + 1} \right){\delta _x}} \right],i \in \left[ {0,{2^L} - 1} \right],
\end{equation}
and discuss the effect of intervals ${I_0}$ and ${I_{{2^L} + 1}}$ separately.

In experiment, the alphabet of each interval $\left\{ {{{\tilde I}_i}} \right\}$ is always mapped to a binary $L$-bits number, namely $X_D^i: = a_L^{\left( i \right)}a_{L - 1}^{\left( i \right)}...a_1^{\left( i \right)}$ with $a_m^{\left( i \right)} \in \left\{ {0,1} \right\}$, which is easy to be processed by the data processing. $a_L^{\left( i \right)}$ and $a_1^{\left( i \right)}$ are the most significant bit (MSB) and the LSB respectively in Gray's binary decomposition \cite{Patent.No.2632058}, and $a_m^{\left( i \right)}a_{m - 1}^{\left( i \right)}...a_1^{\left( i \right)}$ denotes the $m$-LSBs.

The $m$-LSBs operation is a simple postprocessing method which discards several $k$-MSBs in raw data and remains the rest $m$-LSBs, where $m+k=L$, and then outputs the low $m$-bits sequences as the new random bits. This data processing method often shows its effectiveness for the randomness extraction in many scenarios, which can remove the discretization intervals which do not contribute to the final random streams and can uniform the probability distribution of the output data. We give the following examples to show the relations between the raw data and the $m$-LSBs data (only consider data within intervals $\left\{ {{{\tilde I}_i}} \right\}$):

(i) $1$-LSB operation: If we only keep the LSB of the binary raw data $X_D^i$ and discard the other bits, the remaining data is $X_{1 - {\rm LSB}}^i: = a_1^{\left( i \right)}$, which only has two outcomes, 1 or 0. Hence the new intervals $\left\{{\Gamma _j}\right\}$ can be described after recombining the previous intervals, which are given by ($i \in \left[ {0,{2^L} - 1} \right]$):
\begin{align}\
{\Gamma _0} = \left\{ {{{\tilde I}_i}|a_1^{\left( i \right)} = 0} \right\} = \left\{ {{{\tilde I}_i}|i = 0,2, \cdots ,{2^L} - 2} \right\},        \notag\\
{\Gamma _1} = \left\{ {{{\tilde I}_i}|a_1^{\left( i \right)} = 1} \right\} = \left\{ {{{\tilde I}_i}|i = {\rm{1}},{\rm{3}}, \cdots ,{2^L} - 1} \right\}.
\end{align}
The previous intervals where the LSB is equal to zero are combined into a new interval ${\Gamma _0}$ and the previous intervals where the LSB is equal to one are combined into a new interval ${\Gamma _1}$. Hence a part of the POVM operators of the $1$-LSB operation can be defined as (without considering the infinite-length intervals)
\begin{equation}\
\hat M_{1 - {\rm LSB},{\delta _x}}^j = {\left\{ {\int_{{\Gamma _j}} {dx} \left| x \right\rangle \left\langle x \right|} \right\}_{j \in \left[ {0,1} \right]}}.
\end{equation}

(ii) $2$-LSBs operation: If we keep the lowest two bits of the binary raw data $X_D^i$ and cut off the other bits, the remaining data is $X_{2 - {\rm LSBs}}^i: = a_2^{\left( i \right)}a_1^{\left( i \right)}$, and each bit has two outcomes. The new intervals after regrouping the data are given by ($i \in \left[ {0,{2^L} - 1} \right]$):
\begin{align}\
{\Gamma _0} = \left\{ {{{\tilde I}_i}|a_2^{\left( i \right)}a_1^{\left( i \right)} = 00} \right\}
= \left\{ {{{\tilde I}_i}|i \equiv {\rm{0}}\;\left( {\bmod \;4} \right)} \right\},        \notag\\
{\Gamma _1} = \left\{ {{{\tilde I}_i}|a_2^{\left( i \right)}a_1^{\left( i \right)} = 01} \right\}
 = \left\{ {{{\tilde I}_i}|i \equiv {\rm{1}}\;\left( {\bmod \;4} \right)} \right\},       \notag\\
{\Gamma _2} = \left\{ {{{\tilde I}_i}|a_2^{\left( i \right)}a_1^{\left( i \right)} = 10} \right\}
 = \left\{ {{{\tilde I}_i}|i \equiv {\rm{2}}\;\left( {\bmod \;4} \right)} \right\},        \notag\\
{\Gamma _3} = \left\{ {{{\tilde I}_i}|a_2^{\left( i \right)}a_1^{\left( i \right)} = 11} \right\}
= \left\{ {{{\tilde I}_i}|i \equiv {\rm{3}}\;\left( {\bmod \;4} \right)} \right\},
\end{align}
and a part of the POVM operators of the 2-LSBs operation can be written as
\begin{equation}\
\hat M_{2 - {\rm LSBs},{\delta _x}}^j = {\left\{ {\int_{{\Gamma _j}} {dx} \left| x \right\rangle \langle x|} \right\}_{j \in \left[ {0,3} \right]}}.
\end{equation}

(iii) $m$-LSBs operation: Similarly, if we keep the lowest $m$ bits and discard the other bits, the recombing intervals after the $m$-LSBs operation are

\begin{align}\
{\Gamma _0} = \left\{ {{{\tilde I}_i}|a_m^{\left( i \right)}...a_1^{\left( i \right)} = 0...0} \right\}, \notag\\
{\Gamma _{\rm{1}}} = \left\{ {{{\tilde I}_i}|a_m^{\left( i \right)}...a_1^{\left( i \right)} = 0...1} \right\},        \notag\\
.....          \notag\\
{\Gamma _{{2^m} - 2}} = \left\{ {{{\tilde I}_i}|a_m^{\left( i \right)}...a_1^{\left( i \right)} = 1...0} \right\},          \notag\\
{\Gamma _{{2^m} - 1}} = \left\{ {{{\tilde I}_i}|a_m^{\left( i \right)}...a_1^{\left( i \right)} = 1...1} \right\}.
\end{align}

Therefore it is easy to define new discretized intervals of the $m$-LSBs data, given by ($i \in \left[ {0,{2^L} - 1} \right],j \in \left[ {0,{2^m} - 1} \right]$)
\begin{equation}\
{\Gamma _j} = \left\{ {{{\tilde I}_i}|i \equiv j\;\left( {\bmod \;{2^m}} \right)} \right\},
\end{equation}
and a part of the POVM operators of the $m$-LSBs operation can be described, given by
\begin{equation}\
\hat M_{m - {\rm LSBs},{\delta _x}}^j = {\left\{ {\int_{{\Gamma _j}} {dx} \left| x \right\rangle \langle x|} \right\}_{j \in \left[ {0,{2^m} - 1} \right]}}.
\label{LSB_original}
\end{equation}

Considering the infinite-length intervals, the overall POVM operators of $m$-LSBs operation (in the following paper we call them the $m$-LSBs operators) are $\left\{ {\hat M_{m - {\rm LSBs},{\delta _x}}^j,\hat M_{m - {\rm LSBs}}^\infty } \right\}$ $({j \in \left[ {0,{2^m} - 1} \right]})$, where $\hat M_{m - {\rm LSBs}}^\infty  = \mathbb{I} - \sum\nolimits_j {\hat M_{m - {\rm LSBs},{\delta _x}}^j}$ stands for the event on the outside of the detection range. For the convenience of the follow-up discussions, we rewrite Eq.(\ref{LSB_original}) as
\begin{equation}\
\hat M_{m - {\rm LSBs},{\delta _x}}^j = \sum\limits_{i = 0}^{{2^{L \!-\! m}} \!-\! 1} {\int_{\left( {j + {2^m}i} \right){\delta _x}}^{\left( {j + {2^m}i + 1} \right){\delta _x}} {dx} \left| x \right\rangle \left\langle x \right|} ,
\end{equation}
where $j \in \left[ {{\rm{0}},{2^m} - 1} \right]$. Likewise, the $s$-LSBs operators for measuring quadrature $P$ can be written as
\begin{align}\
\hat N_{s - {\rm LSBs},{\delta _p}}^j &= \sum\limits_{i = 0}^{{2^{N \!-\! s}} \!-\! 1} {\int_{\left( {j + {2^s}i} \right){\delta _p}}^{\left( {j + {2^s}i + 1} \right){\delta _p}} {dp} \left| p \right\rangle \langle p|} ,                   \notag\\
\hat N_{s - {\rm LSBs},{\delta _p}}^\infty  &= \mathbb{I} - \sum\nolimits_j {\hat N_{s - {\rm LSBs},{\delta _p}}^j}
\end{align}
where $j \in \left[ {{\rm{0}},{2^s} - 1} \right]$ and $N$ is the number of discretization bits for quadrature $P$.

\section{Trusted source scenario}
\label{Trusted_source_scenario}
In the trusted source scenario, it is always assumed that the source can be well characterized and the technical noises cannot be accessible to the potential adversaries. In this case, one can ensure that the measurement is always within the range $\left[ { - \alpha ,\alpha } \right]$ by adjusting the range of ADCs, and the probabilities of the infinite-length intervals ${I_0}$ and ${I_{{2^L} + 1}}$ (denoted by ${\rm Tr}\left( {\rho \hat M_{m - {\rm LSBs}}^\infty} \right)$) are small, which do not have the contributions for the min-entropy in generating random bits and thus can be neglected. The probabilities of measuring each interval in $\left\{{\Gamma _j}\right\}$ are given by
\begin{equation}\
\Pr \left( {X_{m - {\rm LSBs}}^j} \right) = {\rm Tr}\left( {\rho \hat M_{m - {\rm LSBs},{\delta _x}}^j} \right),
\end{equation}
and the min-entropy is defined as \cite{IEEE.Trans.Inf.Theory.55.4337.2009}
\begin{equation}\
{{H_\infty }}\left( {{x_M}} \right) =  - {\log _2}\left( {{\mathop{\max }_{j \in \left[ {0,{2^m} - 1} \right]}}\Pr \left( {X_{m - {\rm LSBs}}^j} \right)} \right),
\end{equation}
where ${{x_M}}$ stands for the measurement results. The min-entropy corresponds to the maximum guessing probability for an adversary about $X_{m - {\rm LSBs}}^j$, and it is also known as the maximum (almost) uniform randomness which can be extracted out of the distribution $\Pr \left( {X_{m - {\rm LSBs}}^j} \right)$.

In the case of measuring vacuum state with homodyne detection \cite{Nat.Photonics.4.711.2010,Phys.Rev.A81.063814.2010,Appl.Phys.Lett.98.231103.2011}, the outcomes are random and have a probability density function (PDF) ${p_M}\left( {x_M} \right)$, which is Gaussian and centred at zero with variance $\sigma _M^2$, given by
\begin{equation}\
{p_M}\left( {{x_M}} \right) = \frac{1}{{\sqrt {2\pi \sigma _M^2} }}\exp \left( { - \frac{{{{x_M}^2}}}{{2\sigma _M^2}}} \right).
\end{equation}
In practice, the measurement results may not totally originate from the measurements of quantum sources, and some classical noises like the electrical noise will inevitably contribute to the measurement results. To avoid the effects of classical noises and extract randomness independent to classical side information, the influences of classical noises on the minimum entropy need to be characterized \cite{Phys.Rev.A.91.062316.2015,Appl.Phys.Lett.98.231103.2011,Phys.Rev.Appl.3.054004.2015}. The measured signal $M$ is then written as $M=Q+E$ and the resulting measurement PDF ${p_M}$ is a convolution of ${p_Q}$ and ${p_E}$, where ${p_Q}$ and ${p_E}$ denote the PDF of the quantum signal and the classical noises, respectively. To extract randomness independent of classical noises, the conditional PDF between the measured signal and the classical noises need to be considered, which is given by \cite{Appl.Phys.Lett.98.231103.2011,Phys.Rev.Appl.3.054004.2015}
\begin{align}\
{p_{M|E}}\left( {{x_M}|{e}} \right) &= \frac{1}{{\sqrt {2\pi \left( {\sigma _M^2 - \sigma _E^2} \right)} }}\exp \left[ { - \frac{{{{\left( {{x_M} - {e}} \right)}^2}}}{{2\left( {\sigma _M^2 - \sigma _E^2} \right)}}} \right]         \notag\\
&= \frac{1}{{\sqrt {2\pi \sigma _Q^2} }}\exp \left[ { - \frac{{{{\left( {{x_M} - {e}} \right)}^2}}}{{2\sigma _Q^2}}} \right],
\end{align}
where ${\sigma _E^2}$ is the variance of classical noises and $e$ is the influence of classical noises on the measured data. After discretizing the raw data and performing the $m$-LSBs operation, the conditional probability distribution of the final output sequences under the condition of classical noises is the integrals in each intervals $\left\{ {{\Gamma _j}} \right\}$, given by
\begin{equation}\
{{\Pr}_{M|E}}\left( {x_M^j|e} \right) = \int_{{\Gamma _j}} {d{x_m}} {p_{M|E}}\left( {{x_M}|e} \right).
\end{equation}
Therefore, the worst-case min-entropy conditioned on classical side-information $E$ can be defined to describe the lower bound of the randomness of the source, which is given by
\begin{equation}\
{H_\infty }\!\left( {{x_M}|E} \right) \!=\!  \!-\! {\log _2}\!\left( {\mathop {\max }\limits_{e \in R} \mathop {\max }\limits_{j \in \left[ {0,{2^m}\! -\! 1} \right]} \! {\Pr }_{M|E} \!\left( {x_M^j|e} \right)}\! \right).
\end{equation}
Here we assume that the classical noises are well-characterized and independent to the quantum signal \cite{Phys.Rev.A.91.062316.2015,Phys.Rev.Appl.3.054004.2015}, which is reasonable for the trusted source scenario. For practical purposes, we bound the maximum excursion of $e$ as $- 12{\sigma _E} \le e \le 12{\sigma _E}$, which is valid with the probability greater than $99.99\%$.

\begin{figure}[b]
	\centering
	\includegraphics[width=8cm]{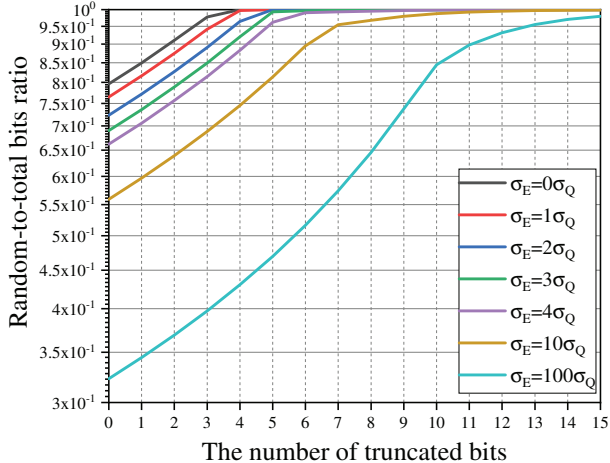}\\
	\caption{(Color online) The random-to-total bits ratio vs. the number of truncated bits after using the $m$-LSBs postprocessing method under the trusted source assumption. Different classical noise levels are considered. From top to bottom, the standard deviations of classical noises are 0, 1, 2, 3, 4, 10 and 100 times than that of the quantum signals.}
	\label{RTBR}
\end{figure}

We exploit the \emph{random-to-total bits ratio} (RTBR) ${{R = {H_{\infty }}\left( {{x_M}|E} \right)} \mathord{\left/{\vphantom {{R = {H_{\infty }}\left( {{x_M}|E} \right)} m}} \right.
 \kern-\nulldelimiterspace} m}$ to quantify the randomness of the final sequences after the $m$-LSBs operation. Noting that we can also define the statistical
distance between two probability distributions $X$ and $Y$ to measure the distance of final data's distribution from the uniformly distributed numbers \cite{Rev.Mod.Phys.89.015004.2017}, but we think the RTBR is more intuitive in the $m$-LSBs scenarios. If $R = 1$, it means that the min-entropy is equal to the Shannon entropy and all the remaining bits can be used for generating random numbers.

Supposing that the ADC has 16-bits resolution, we plot the relations between the number of truncated bits and the RTBRs under different classical noise levels, which are shown in Fig.\ref{RTBR}. It can be seen that the min-entropy cannot reach the maximum value (corresponding to the RTBR closing to $1$) without postprocessing, and even there is no classical noises. This is the direct evidence that the random number is not a uniform distribution after the measurement step. Intuitively, the simple postprocessing method like the $m$-LSBs operation can ``uniform'' measurement results. We should note that this kind of postprocessing method cannot raise the total randomness in the entropy source, but can only increase the proportion of random bits after processing, and ``worse'' sources mean ``more'' truncated bits are needed to extract more randomness. Moreover, even though the classical noise is much larger than the quantum signal (${\sigma _E} \sim {\rm{100}}{\sigma _Q}$), the RTBR can still close to $1$ after discarding several bits. This postprocessing method is very practical in the case of the trusted source scenario, since we do not need to exploit the extractors like Toeplitz-hashing extractor to extract randomness, which is complicated especially in real-time hardware implementations.

\begin{figure}[b]
	\centering
	\includegraphics[width=8cm]{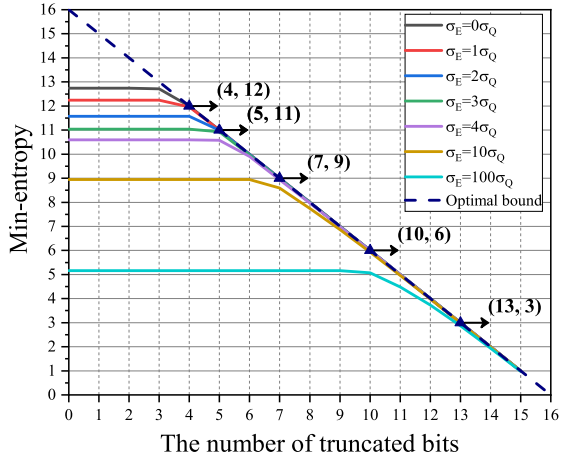}\\
	\caption{(Color online) The min-entropy of the sequence vs. the number of truncated bits after using the $m$-LSBs postprocessing method under the trusted source assumption. Different classical noise levels are considered. From top to bottom, the standard deviations of classical noises are 0, 1, 2, 3, 4, 10 and 100 times than that of the quantum signals. The dashed line is the optimal bound, where the point satisfies that after discarding some bits, the number of remaining bits equals the minimum entropy of the random sequence.}
	\label{min_entropy}
\end{figure}

It should note that high entropy is very important for real implementations which can guarantee both speed and security of QRNGs, and the enhance of extractable randomness has been detailed analyzed in Ref.~\cite{Entropy.20.819.2018}. Different from the randomness extraction method as shown in Ref.~\cite{Entropy.20.819.2018}, the $m$-LSB method in our work cannot enhance the min-entropy in the source, but only makes the min-entropy of the post-processing random sequence closer to the Shannon entropy, that is, making the random sequence tends to the uniform distribution after postprocessing. For experimental purpose, we need to determine the optimal number of the truncated bits for different entropy sources. We plot the relation between the min-entropy of the random sequences and the number of truncated bits after postprocessing, as shown in Fig.~\ref{min_entropy}.

The point of the dashed line in Fig.~\ref{min_entropy} depicts that the number of remaining bits is equal to the min-entropy of the random sequence after discarding several bits, which means all the remaining bits can be used to extract randomness, thus we call this dashed line the ``optimal bound''. The solid lines show the relations between the min-entropy and the number of truncated bits under different classical noise levels. From top to bottom, the standard deviations of classical noises are 0, 1, 2, 3, 4, 10 and 100 times than that of the quantum signals. It can be seen that, when the number of discarded bits is small, the $m$-LSB operation does not reduce the min-entropy of the extracted sequence, which means that discarding few bits is not helpful for randomness extraction. Then if we continue to increase the number of truncated bits, we find that although the min-entropy decreases, the proportion of extractable randomness increases, until the point of intersection with the optimal bound is reached. The intersection points between the optimal bound and the solid lines are the cases where all the remaining data can be exploited to extract random sequences. After the intersection point, increasing the number of truncated bits will only reduce the min-entropy, thus it is not effective for randomness extraction. Therefore, in experiments, we usually choose the intersection points between the curves of min-entropy changing with the number of truncated bits and the optimal bound as the number of discarded bits. We also mark the intersection points in Fig.~\ref{min_entropy}, where the abscissas represent the number of truncated bits and the ordinates represent the min-entropy of the corresponding points.

\section{Untrusted source scenario}
\label{Untrusted_source_scenario}

In the untrusted source scenario, it is always assumed that the source devices are not perfect and ill-characterized, and all the technical noises have to be attributed to the malicious adversaries and thus cannot be trusted \cite{npj.Quantum.Information.2.16021.2016}. Hence, the quantum side information $E$ for guessing the random sequences ${X_{m - {\rm LSBs}}}$ needs to be considered after the $m$-LSBs operation, which could be quantified by the conditional min-entropy ${H_{\infty }}\left( {{X_{m - {\rm LSBs}}}|E} \right)$.

The EUR is an effective tool for the conditional min-entropy estimation considered in the SDI-QRNG scenarios \cite{Phys.Rev.X.6.011020.2016,Phy.Rev.Lett.118.060503.2017}, assuming with trusted measurement devices and a totally untrusted source. In this section, we construct a new continuous variable EUR with the $m$-LSBs operators to bound the conditional quantum min-entropy by randomly switching between two measurement bases.

\begin{figure}[b]
  \centering
  \includegraphics[width=\linewidth]{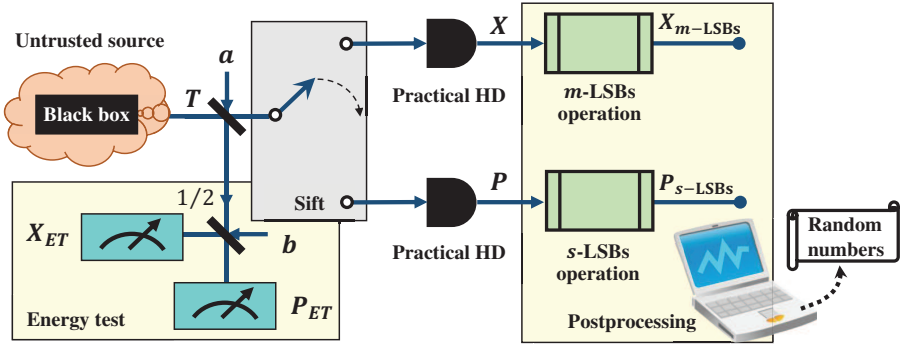}\\
  \caption{(Color online) The structure of the SDI-QRNG with the $m$-LSBs operations. An energy test is performed to bound the energy of modes ${X_{ET}}$ and ${P_{ET}}$ by a heterodyne detector, which are splitted from the source by a beam splitter with transmittance $T$. $a$ and $b$ are two vacuum states induced by beam splitters. HD denotes the homodyne detector.}
  \label{entropic_uncertainty_relation}
\end{figure}

 The SDI-QRNG scenario with the $m$-LSBs operations is shown in Fig.\ref{entropic_uncertainty_relation}. One (Alice) in the lab holds an untrusted and uncharacterized source treated as a black box, where a potential adversary (Eve) may has access to a quantum system $E$ correlated with the source. Since the intervals ${I_0}$ and ${I_{{2^L} + 1}}$ are with infinite length, any uncertainty relations will get trivial bounds within these intervals.

To solve this problem, we use the energy test $\mathcal{T}\left( {M,T} \right)$~\cite{Phys.Rev.A.90.042325.2014,Phys.Rev.A.98.012314.2018} to check whether the probabilities of these two intervals are small, in order to ensure that the purified distance between ${\rho _{\tilde XE}}$ and ${\rho _{XE}}$ is small enough, where $\tilde X$ can be obtained by measuring the amplitudes or phases of the untrusted states according to intervals $\left\{ {{{\tilde I}_k}} \right\}$. Alice uses an unbalanced beam splitter with transmissivity $T$ to split the source into two path. The transmission path is exploited for generating random numbers and the reflection path is used for monitoring the energy of the incoming signals. A heterodyne detection is employed to check if $\left| {{X_{ET}}} \right|,\left| {{P_{ET}}} \right| \le M$ is satisfied for all of the input signals and aborts the QRNG process otherwise. The upper bound of the probability that Alice measures with homodyne detection larger than $\alpha$ can be bounded by the function $\Gamma \left( {\alpha,T,M } \right)$, given by
\begin{equation}\
\Gamma \!\left( {\alpha,T,M } \right): =\! \frac{{\sqrt {1 \!+\! \lambda } \! +\! \sqrt {1\! +\! {\lambda \!^{ - 1}}} }}{2}\exp \!\left( { - \frac{{{{\left( {\mu \alpha \! -\! M} \right)}^2}}}{{T\left( {1\! +\! \lambda } \right)/2}}} \right),
\end{equation}
where $\mu  = \sqrt {\frac{{1 - T}}{{2T}}}$ and $\lambda  = {\left( {\frac{{2T - 1}}{T}} \right)^2}$. The smoothness of the energy test ${\tilde \epsilon }$ further can be written as
\begin{equation}\
\tilde \epsilon  = \sqrt {\frac{{2n\Gamma \left( {\alpha,T,M } \right)}}{{{p_{pass}}}}}.
\end{equation}

Alice then randomly measures the quadrature $X$ or the quadrature $P$ of the incoming signals with practical homodyne detection in many runs. The practical homodyne detector in this scenario is assumed to be trusted modeled by an ideal homodyne detector followed by a practical ADC. After Alice gets enough data, the protocol turns to the postprocessing process. The $m$-LSBs operation for $X$ resulting data ${{X_{m - {\rm LSBs}}}}$ is used for random bits generation and the $s$-LSBs operation for $P$ resulting data ${{P_{s - {\rm LSBs}}}}$ is exploited for the entropy estimation. The smooth min-entropy $H_{\infty }^\epsilon \left( {{X_{m - {\rm LSBs}}}|E} \right)$ can be bounded with the help of the EUR, which is given by
\begin{equation}\
H_{\infty }^\epsilon \!\left( {{X_{m - {\rm LSBs}}}|E} \right) \!\ge \! - {\log _2}\left( {{c_{{\rm LSBs}}}} \right)\! -\! {H_{\max }}\left( {{P_{s - {\rm LSBs}}}} \right),
\end{equation}
where $\epsilon  \le {{\left( {{\epsilon _1} - {\epsilon _s}} \right)} \mathord{\left/
 {\vphantom {{\left( {{\epsilon _1} - {\epsilon _s}} \right)} {\left( {2{p_{pass}}} \right) - 2\tilde \epsilon }}} \right. \kern-\nulldelimiterspace} {\left( {2{p_{pass}}} \right) - 2\tilde \epsilon }}$ \cite{Phys.Rev.Lett.109.100502.2012}, and ${c_{{\rm LSBs}}}$ denotes the maximum overlap between operators $\hat M_{m - {\rm LSBs},{\delta _x}}^j$ and $\hat N_{s - {\rm LSBs},{\delta _p}}^j$, which reads
 \begin{equation}\
{c_{{\rm LSBs}}} = \mathop {\sup }\limits_{i,j} {\left\| {\sqrt {\hat M_{m - {\rm LSBs},{\delta _x}}^i} \sqrt {\hat N_{s - {\rm LSBs},{\delta _p}}^j} } \right\|^2},
\label{overlap_LSB}
\end{equation}
where $\left\|  \cdot  \right\|$ denotes the operator norm (i.e., the maximal singular value). For simplicity, we first consider the overlap term of the differential entropy, which satisfies
\begin{align}\
{c_\infty } &= \mathop {\lim \inf }\limits_{{\delta _x},{\delta _p} \to 0} \left[ {\frac{{{c_{{\rm LSBs}}}}}{{{\delta _x}{\delta _p}}}} \right]             \notag \\
&= \mathop {\sup }\limits_{i,j} \mathop {\lim \inf }\limits_{{\delta _x},{\delta _p} \to 0} \left[ {\frac{1}{{{\delta _x}{\delta _p}}}{{\left\| {\sqrt {\hat M_{m - {\rm LSBs},{\delta _x}}^i} \sqrt {\hat N_{s - {\rm LSBs},{\delta _p}}^j} } \right\|}^2}} \right]                        \notag \\
&= \mathop {\sup }\limits_{i,j} \mathop {\lim \inf }\limits_{{\delta _x},{\delta _p} \to 0} {\left\| {\sqrt {\frac{{\hat M_{m - {\rm LSBs},{\delta _x}}^i}}{{{\delta _x}}}} \sqrt {\frac{{\hat N_{s - {\rm LSBs},{\delta _p}}^j}}{{{\delta _p}}}} } \right\|^2}                                 \notag \\
&= \mathop {\sup }\limits_{i,j} {\left\| {\hat M_{m - {\rm LSBs},\infty }^i\hat N_{s - {\rm LSBs},\infty }^j} \right\|^2},
\end{align}
where ($i \in \left[ {0,{2^m} - 1} \right]$, $j \in \left[ {0,{2^s} - 1} \right]$)
\begin{align}\
\hat M_{m - {\rm LSBs},\infty }^i &= \mathop {\lim }\limits_{{\delta _x} \to 0} \left( {\frac{1}{{{\delta _x}}}\hat M_{m - {\rm LSBs},{\delta _x}}^i} \right)                      \notag \\
&= \sum\limits_{k = 0}^{{2^{L - m}} - 1} {{{\left| {\left( {i + {2^m}k} \right){\delta _x}} \right\rangle }_x}{{\left\langle {\left( {i + {2^m}k} \right){\delta _x}} \right|}_x}}         \notag \\
\hat N_{s - {\rm LSBs},\infty }^j &= \mathop {\lim }\limits_{{\delta _p} \to 0} \left( {\frac{1}{{{\delta _p}}}\hat N_{s - {\rm LSBs},{\delta _p}}^j} \right)                  \notag \\
&= \sum\limits_{k = 0}^{{2^{N - s}} - 1} {{{\left| {\left( {j + {2^s}k} \right){\delta _p}} \right\rangle }_p}{{\left\langle {\left( {j + {2^s}k} \right){\delta _p}} \right|}_p}}.
\end{align}
The overlap of the two LSBs operators in Eq.(\ref{overlap_LSB}) can be well approximated by the relation ${c_{{\rm LSBs}}} \approx {c_\infty } \cdot {\delta _x} \cdot {\delta _p}$.

To bound the overlap term ${c_\infty }$, we first define
\begin{align}\
{U_{j,k}}: &= \left\| {\hat M_{m - {\rm LSBs},\infty }^j\hat N_{s - {\rm LSBs},\infty }^k} \right\|          \notag\\
&=\mathop {\max }\limits_\psi  \left( {\left\langle \psi  \right|\hat M_{m - {\rm LSBs},\infty }^j\hat N_{s - {\rm LSBs},\infty }^k\left| \psi  \right\rangle } \right),
\end{align}
where $\psi$ is an arbitrary wave function. Because the modular square of arbitrary wave functions is not greater than 1, we can easily bound ${U_{j,k}}$ by
\begin{equation}\
\left| {{U_{j,k}}} \right| \!\le \!\sum\limits_{g = 0}^{{2^{L \!-\! m}} \!-\! 1} {\sum\limits_{t = 0}^{{2^{N \!-\! s}} \!-\! 1} {\frac{1}{{\sqrt {2\pi \hbar } }}\exp \left( {\frac{{i\left( {j \!+\! {2^m}g} \right)\left( {k \!+\! {2^s}t} \right){\delta _x}{\delta _p}}}{\hbar }} \right)} },
\end{equation}
and then ${c_\infty } = \mathop {\sup }\limits_{j,k} {\left| {{U_{j,k}}} \right|^2}$ can be obtained.

\begin{figure}[t]
  \centering
  \includegraphics[width=8cm]{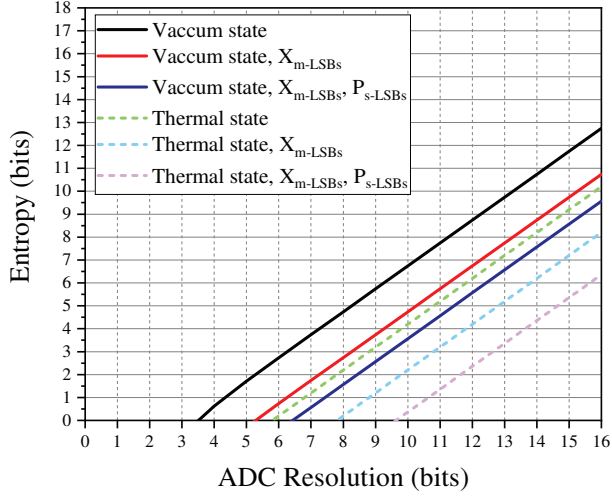}\\
  \caption{(Color online) Entropy evaluation results in SDI-QRNG with LSBs operation. We show the entropy extraction both in vacuum-state (solid lines) and thermal-state (dashed lines) source under different ADC Resolutions. ${X_{m - {\rm LSBs}}}$ and ${P_{s - {\rm LSBs}}}$ denote that $m - {\rm LSBs}$ and $s - {\rm LSBs}$ operations are used for processing of data $X$ and $P$ respectively. Both $m$ and $s$ are chosen as ADC resolution minus one bit.}
  \label{SDI_results}
\end{figure}

We plot the relations between extracted random bits (entropy) and the resolutions of ADCs in the SDI-QRNG scenario as shown in Fig.\ref{SDI_results}. We only consider the asymptotic regime in our paper, for the finite-size effect can be easily introduced by the smooth min-entropy, which has been already discussed in Ref.\cite{Phy.Rev.Lett.118.060503.2017}. Both vacuum states (solid lines) and thermal states (dashed lines) as the randomness sources are displayed. We discuss the performances of the SDI-QRNGs with three different $m$-LSBs postprocessing methods, which can be applied to all entropy sources: (1) the most common SDI-QRNG condition, which has been discussed in Ref.\cite{Phy.Rev.Lett.118.060503.2017}, and no $m$-LSBs operation is used for processing data $X$ and data $P$. (2) the $m$-LSBs operation is performed for processing data $X$ ($X$ is used for generating random bits in the SDI-QRNG) and has no operation for data $P$ ($P$ is used for the entropy estimation in the SDI-QRNG). (3) the  $m$-LSBs operation is performed for processing data $X$ and the $s$-LSBs operation is used for operating data $P$. Both $m$-LSBs and $s$-LSBs operations are chosen to discard the MSB of the data and keep the rest bits in our discussions. It is shown that, whether the LSBs operations are performed on the checking data or the raw random numbers, discarding one per bit of data will lose two bits of the final secure random numbers.

It is intuitive that the $m$-LSBs operation has no contribution to distill quantum randomness in the SDI-QRNG scenario. The reason is that the EUR is a relative tight bound for estimating randomness of the entropy sources, which can extract secure random bits out of the potential quantum side information without extra classical postprocessing operations.

\section{Discussion}
\label{Discussion}

Let's investigate why such deterministic operations like the $m$-LSBs operation are not helpful for the randomness extraction when there exists quantum side information.

Assuming $X:{\rm{ = }}{\left\{ {{\rm{0,1}}} \right\}^L}$ is the random number space before the post-processing perfermed, and the output random number $x$ takes values in this set. Considering that $X$ may be correlated to a quantum system $E$, which is accessible to a potential Eve, and for every possible outcome $X=x$, Eve holds a state $\rho _E^x$. The joint state of $X$ and $E$ can then be described as a classical-quantum state (cq-state), which is given by
\begin{equation}\
{\rho _{XE}} = \sum\limits_{x \in X} {{P_X}\left( x \right)\left| x \right\rangle \left\langle x \right| \otimes \rho _E^x},
\end{equation}
where we use the orthogonal basis ${\left\{ {\left| x \right\rangle } \right\}_{x \in X}}$ to represent the family of classical random numbers $x$, and ${P_X}\left( x \right)$ is the probability that $X=x$. The maximum probability of guessing $X$ given $E$ can be defined by ${p_{guess}}\left( {{\rho _{XE}}|E} \right)$, namely,
\begin{equation}\
{p_{guess}}\left( {{\rho _{XE}}|E} \right): = \mathop {\sup }\limits_{\left\{ {{M_x}} \right\}} \sum\limits_{x \in X} {{P_X}\left( x \right)Tr} \left( {\rho _E^x{M_x}} \right),
\label{p_guess_x}
\end{equation}
where the supreme is taken over all the POVMs ${\left\{ {{M_x}} \right\}_{x \in X}}$ on $E$, and the min-entropy condition on Eve's auxiliary system is defined as
\begin{equation}\
{H_{\min }}\left( {{\rho _{XE}}|E} \right) =  - {\log _2}{p_{guess}}\left( {{\rho _{XE}}|E} \right).
\end{equation}

Now if Alice processes her random numbers by a mapping-type operation, which can map data from a large data set to a smaller one, this is equivalent to apply a deterministic classical operation $\hat N$ to the cq-state, namely, $\hat N = {\hat N_x} \otimes {I_E}$. More specifically, the operation $\hat N$ is acting on the classical part of the cq-state and cannot change the quantum part held by eavesdroppers. As mentioned above, such a ``mapping" operation is a selective combination of the discrete intervals. For instance, if Alice exploits the $m$-LSBs operation to perform randomness extraction, the random number space after postprocessing becomes $Y:{\rm{ = }}{\left\{ {{\rm{0,1}}} \right\}^m}$. Assuming that the classical values $y{ \in Y}$ are the outcomes after post-processing, which are denoted by the orthogonal basis ${\left\{ {\left| y \right\rangle } \right\}_{y \in Y}}$. The cq-state ${\rho _{YE}}$ after the postprocessing process should satisfy
\begin{equation}\
{\rho _{YE}} = \sum\limits_{y \in Y} {{P_Y}\left( y \right)\left| y \right\rangle \left\langle y \right| \otimes \rho _E^y}.
\end{equation}
This is because the classical operation $\hat N$ Alice exploited will not affect Eve's hidden systems, and Eve will always get the quantum state $\rho _E^y$ by recombining some of his original auxiliary systems $\rho _E^x$. Here we assume that Eve fully knows the deterministic operation performed on Alice's side.

Then it is important to know the guessing probability ${p_{guess}}\left( {{\rho _{YE}}|E} \right)$ after post-processing process, which is given by
\begin{equation}\
{p_{guess}}\left( {{\rho _{YE}}|E} \right): = \mathop {\sup }\limits_{\left\{ {{M_y}} \right\}} \sum\limits_{y \in Y} {{P_Y}\left( y \right)Tr} \left( {\rho _E^y{M_y}} \right),
\label{p_guess_y}
\end{equation}
where the supreme is taken over all the POVMs ${\left\{ {{M_y}} \right\}_{y \in Y}}$ on $E$. Noticing that the guessing probability in Eq.(\ref{p_guess_y}) differs from that before postprocessing in two places. One is the classical probabilities ${P_Y}\left( y \right)$ and ${P_X}\left( x \right)$, and the other is the probabilities that Eve implements the optimal measurements to get the results $y$ or $x$, which are denoted by $Tr\left( {\rho _E^y{M_y}} \right)$ and $Tr\left( {\rho _E^x{M_x}} \right)$, respectively. The classical probability ${P_Y}\left( y \right)$ is the summation of some of the probabilities ${P_X}\left( x \right)$ by recombining the measurement intervals, which means that the probability ${P_Y}\left( y \right)$ is greater than ${P_X}\left( x \right)$. The term $Tr\left( {\rho _E^y{M_y}} \right)$ in Eq.(\ref{p_guess_y}) is not smaller than the original term $Tr\left( {\rho _E^x{M_x}} \right)$ in Eq.(\ref{p_guess_x}) owning to the fact that Eve can also get more knowledge by measuring auxiliary systems $\rho _E^y$ rather than $\rho _E^x$.

Consequently, it is easy to prove that the relation
\begin{equation}\
{p_{guess}}\left( {{\rho _{YE}}|E} \right) \ge {p_{guess}}\left( {{\rho _{XE}}|E} \right)
\end{equation}
always holds and the min-entropy condition on $E$ will not be increased after the deterministic ``mapping" operation.

\section{Conclusion}
\label{Conclusion}
We have presented an analysis of the $m$-LSBs postprocessing operation in QRNGs. The new $m$-LSBs operators have been constructed by regrouping the coarse-grained measurement intervals. We also applied the $m$-LSBs operators to both trusted source and untrusted source scenarios. The trusted source scenario was analyzed by the worst-case min-entropy conditioned on classical side-information, while the EUR was exploited to study the untrusted source scenario. We find that the $m$-LSBs operation can be exploited as an effective randomness extractor in the case of trusted sources, but not in the case of untrusted sources. We believe that the $m$-LSBs operation can be used as a candidate of randomness extractors for real-time hardware implementations, especially in the trusted-device scenarios.

We note that our method of constructing the $m$-LSBs operators can also be applied to the other types of simple postprocessing methods, such as exclusive-OR operations and the von Neumann debiasing method. The key point is to find the ``data mapping table'' before and after data processing. An interesting extension of this paper would be to further study other simple postprocessing methods by suitable recombining the coarse-grained measurement intervals.

\section*{Acknowledgments}
This work is supported by the National Natural Science Foundation under Grant (Grant No. 61531003, 61771439), the National Science Fund for Distinguished Young Scholars of China (Grant No. 61225003), and China Postdoctoral Science Foundation (Grant No. 2018M630116).




\end{document}